\RequirePackage{fixltx2e}

\documentclass[twocolumn,showpacs,preprintnumbers,amssymb]{revtex4}

\usepackage{graphics}
\usepackage{epsfig} 
\usepackage{hyperref}
\usepackage{graphics}
\usepackage{color}      
\usepackage{graphicx}
\usepackage{graphicx}
\usepackage{dcolumn}

\begin{document}

\title { Entropy production and entropy extraction rates for a Brownian  particle that walks in underdamped medium}
\author{Mesfin Asfaw  Taye}
\affiliation {West Los Angles College, Science Division \\9000  Overland Ave, Culver City, CA 90230, USA}

\email{tayem@wlac.edu}

\begin{abstract}
The  expressions for   entropy production,    free energy,  and    entropy extraction rates  are derived for a Brownian  particle that walks in an underdamped medium. Our analysis indicates that    as long as the system is  driven out of equilibrium, it constantly produces entropy at the same time it extracts  entropy out of the system.  At steady state,  the rate of entropy production ${\dot e}_{p}$ balances the rate of entropy extraction ${\dot h}_{d}$. At equilibrium both entropy production and extraction rates become zero. The 
entropy production and  entropy extraction rates  are also sensitive to time. As time progresses,  both entropy production and extraction rates increase in time and saturate to  constant values. Moreover employing microscopic stochastic approach, several  thermodynamic   relations for different model systems are explored analytically and via numerical simulations by considering a Brownian particle that moves in overdamped medium. Our analysis indicates that  the results obtained for underdamped cases  quantitatively agree with  overdamped cases at steady state. The fluctuation theorem is also  discussed in detailed.  

\end{abstract}
\pacs{Valid PACS appear here}
\maketitle



 \section{Introduction}

Exploring the   thermodynamic feature of equilibrium systems  is vital and recently   have received  significant attentions  since these systems  serve as a starting point to study the thermodynamic properties of systems which are far from equilibrium. Because  most physically relevant systems are far from equilibrium, it is vital to explore the thermodynamic properties of systems which are driven out of equilibrium. However such systems are often challenging  since their thermodynamic relations  such as entropy and free energy depend on their reaction rates. Despite the challenge,  the thermodynamic relations of systems which are far from equilibrium  are explored in the works \cite{mu1,mu2,mu3,mu4}. Particularly, the  
Boltzmann-Gibbs nonequilibrium entropy
along with the entropy balance equation serves as an important tool to explore 
the nonequilibrium thermodynamic features \cite{mu1,mu2,mu3}. 

In the past,  microscopic stochastic approach has been  used by 
Schnakenberg to derive  various thermodynamic quantities   such as entropy production rate in terms of local probability density and transition probability rate \cite{mu3}. Later, many theoretical studies were conducted see for example the works 
\cite{mu4,mu5,mu6,mu7,mu8,mu9,mu10,mu11,mu12,mu13,mu14,mu15,mu16}. Recently,  we presented an exactly solvable  model and studied  the factors that affect the entropy production and extraction rates  \cite{mu17,muu17,muuu17} for a Brownian particle that walks on discrete lattice system. More recently, using Boltzmann-Gibbs nonequilibrium entropy, we derived the general expressions for the free energy, entropy production and    entropy extraction rates for a Brownian particle moving in a viscous medium where the dynamics of its motion is governed by the Langevin equation. Employing   Boltzmann-Gibbs nonequilibrium entropy as well as  from  the knowledge of 
local probability density and particle current, it is shown that as long as the system is far from equilibrium, it constantly produces entropy  
and at the same time extracts entropy out of the system. 
Since  many biological  problems such as intracellular transport of kinesin or dynein inside the cell can be studied by considering a simplified model of particles walking on lattice as discussed  in works  by   
T. Bameta $et$. $al$. \cite{mu28}, D. Oriola $et$. $al$.  \cite{mu29} and  O. Campas $et$. $al$. 
\cite{mu30}, the model considered  will serve as a starting point to study the thermodynamics features of two or more interacting particles hopping on a lattice. At this point, it is important to stress that most of our studies are  focused on exploring the  thermodynamic property of systems that operate in the  classical regimes. For systems that operate at quantum realm,   the dependence of  thermodynamic quantities on the model parameters is studied in the works  \cite{mu25,mu26,mu27}.  Particularly, Boukobza. $et$. $al.$ investigated  the thermodynamic feature of a three-level maser. Not only  the entropy production rate is defined  in terms   of  the system  parameters but it is shown that the first and second laws of thermodynamics are always satisfied in the model  system \cite{mu27}.

In this work, using Langevin equation and  Boltzmann-Gibbs nonequilibrium entropy, the general expressions for the free energy, entropy production  ${\dot e}_{p}$  and    entropy extraction rates  ${\dot h}_{d}$  are derived  in terms of velocity and probability distribution considering a Brownian particle  that moves in  underdamped medium. It is shown that the  entropy production and extraction rates increase in time and saturate to a constant value. At steady state, the rate of entropy production balances the rate of entropy extraction while at equilibrium both entropy production and extraction rates become zero.  Moreover, after extending  the results obtained by Tome. $et.$ $al.$ \cite{ta1} to a spatially varying temperature case, we further analyze  our model systems. Once again,  we show that 
the entropy production rate ${\dot e}_{p}$ increases  in time and at steady state (in the presence of load), ${\dot e}_{p}={\dot h}_{d}>0$. At stationary state (in the absence of load), ${\dot e}_{p} = {\dot h}_{d}=0$. Moreover, 
when the particle hops in nonisothermal medium where the medium temperature linearly  decreasing  (in the presence of load),  the exact analytic results exhibit that  the velocity approach zero when the load approach zero and as long as a periodic boundary condition is imposed. We also show that the approximation performed based on Tome. $et.$ $al.$ \cite{ta1} and our   general analytic  expression agree quantitively. The analytic results also justified via numerical  simulations.
 
Furthermore,  we   discuss the non-equilibrium thermodynamic features of   a Brownian particle that  hops in a ratchet potential where the potential is coupled with a spatially  varying temperature. It is shown that the operational regime  of such Brownian heat engine is dictated by the magnitude of the external load $f$. The steady state current or equivalently the velocity of the engine is positive   when $f$ is smaller and the engine acts as a heat engine. In this regime ${\dot e}_{p}={\dot h}_{d}>0$.  When $f$ increases, the velocity of the particle decreases and at stall force, we find that ${\dot e}_{p}={\dot h}_{d}=0$ showing that the system is reversible  at this particular choice of parameter.  For large load, the current is negative and the engine acts as a refrigerator. In this region ${\dot e}_{p}={\dot h}_{d}>0$. Here we first  study the underdamped case via simulations and then for overdamped case, the thermodynamic feature for the model system is explored analytically.

The rest of paper is organized as follows: in Section II, we present the model system as well as the  derivation of entropy production and free energy.  In Section III, we   explore the dependence  for the  entropy production, entropy  exaction  and  free energy rates on the system parameters for a Brownian particle that freely  diffuses  in isothermal underdamped medium.  In section IV,  the dependence  for various thermodynamic quantities  on system parameters   is explored considering a Brownian particle that undergoes a biased random walk in
a spatially varying thermal arrangement in the presence of external load.   In section V, we consider a Brownian particle walking in rachet potential. The fluctuation theorem is discussed in section VI. Section VII deals with summary and conclusion.

\section{Free energy and Entropy production }  

In the work  \cite{ta1},  the expressions for  entropy production and entropy extraction rates  were presented in terms of  particle velocity and probability distribution   considering  underdamped and isothermal medium.  For  a spatially varying  thermal arrangement, next we derive the  thermodynamic quantities by  considering  a single Brownian  particle that  hops in underdamped medium along the   potential  $U(x)=U_{s}(x)+fx$ where  $U_{s}(x)$ and $f$ are the periodic  potential  and the external force, respectively.

For a single particle that is arranged to undergo a random walk, the dynamics of the particle is
governed  by  Langevin equation
\begin{equation}
m{dv\over dt} = -\gamma{dx\over dt}- {d U(x) \over dx}  + \sqrt{2k_{B}\gamma T(x)}\xi(t).
\end{equation}
The  Boltzmann constant  $k_{B}$  is assumed to be unity. The random noise $\xi(t)$ is assumed to be Gaussian white noise satisfying the relations 
$\left\langle  \xi(t) \right\rangle =0$ and $\left\langle \xi(t)  \xi(t') \right\rangle=\delta(t-t')$.  The viscous  friction  $\gamma$  is assumed to be constant while the temperature $T(x)$ varies along the medium.
For underdamped Langevin case neither Ito nor  Stratonovich interpretation is needed  as discussed   by Sancho. $et$ .$at$  \cite{am3}  and Jayannavar   $et$ .$at$  \cite{am33}.

For overdamped case,  the above Langevin equation can be written as 
\begin{eqnarray}
\gamma(x){dx\over dt}&=&{-\partial U(x)\over \partial x} -{(1-\epsilon)\over \gamma(x)}{\partial\over   \partial x}(\gamma(x)T(x))+ \nonumber \\
&&\sqrt{2k_{B}\gamma(x) T(x)}\xi(t).
\end{eqnarray}
Here the  Ito  and  Stratonovich interpretations correspond to the case where $\epsilon=1$  and $\epsilon=1/2$, respectively while the case  $\epsilon=0$ is   called   the  H\"anggi   a post-point or transform-form interpretation \cite{am1,am2, am3}. 

 The   Fokker-Plank equation for underdamped case  is given by
\begin{eqnarray}
{\partial P\over \partial t}&=&-{\partial (vP) \over \partial x}-{1 \over m}{\partial(U'(x)P) \over \partial v}+ \nonumber \\
&&{\gamma \over m}{\partial (vP) \over \partial v}+{\gamma T(x) \over m^2}{\partial^2 P \over \partial v^2}
\end{eqnarray}
where $P(x,v,t)$ is the probability  of finding the particle at particular position, velocity and time.
The   Gibbs entropy is given by 
\begin{eqnarray}
S(t)= -\int P(x,v,t)\ln P(x,v,t) dxdv.
\end{eqnarray}
The entropy production  and dissipation rates  can be derived  via the  approach stated in the work \cite{mu7}. The  derivative of $S$  with time  leads to
\begin{eqnarray}
{d S(t)\over dt}&=& -k_{B}\int {\partial P(x,v, t)\over \partial t} \ln[P(x,v,t)]dxdv.
\end{eqnarray}
Eq. (5) can be rewritten as  
\begin{eqnarray}
{d S(t)\over dt}&=&{\dot e}_{p}-{\dot h}_{d} 
\end{eqnarray}
where  ${\dot e}_{p}$ and   ${\dot h}_{d}$ are the entropy production   and  extraction rates.

In order to calculate ${\dot h}_{d}$, let us first  find the heat dissipation rate ${\dot H}_{d}$   via stochastic energetics that discussed in the works \cite{am4,am5}. Accordingly  the energy  extraction rate can be written as 
\begin{eqnarray}
{\dot H}_{d}  
&=&-\left\langle \left(-\gamma(x){\dot x}+ \sqrt{2k_{B}\gamma(x) T(x)}\right).{\dot x}\right\rangle  \nonumber \\
&=&-\left\langle m{vdv\over dt}  +v U'(x)   \right\rangle.
\end{eqnarray}
Once the energy dissipation rate is obtained,  
based on our previous works \cite{mu17,muu17,muuu17},    the entropy extraction rate ${\dot h}_{d}$ then can be found as 
\begin{eqnarray}
{\dot h}_{d}  
&=&-\int \left({m{vdv\over dt}  +v U'(x) \over T(x)} \right)P dxdv.
\end{eqnarray}
At this point  we want to stress that Eq. (8) is exact and do not depend on any boundary condition  (as it can be seen in  the next sections).
Since ${d S(t)\over dt}$ and ${\dot h}_{d}  $ are computable,  the entropy production rate  can be readily  obtained as 
\begin{eqnarray}
{\dot e}_{p}&=&{d S(t)\over dt}+{\dot h}_{d}. 
\end{eqnarray}

In high friction limit, Eq. (8)  converges to
\begin{eqnarray}
{\dot h}_{d} &=& -\int \left[ J{ U'(x)\over T(x)}\right]dx
\end{eqnarray} 
 where the probability current 
\begin{eqnarray}
J(x,t)&=&-\left[U'(x)P(x,t) +T(x){\partial P(x,t) \over \partial x}\right].
\end{eqnarray}

At steady state ${d S(t)\over dt}=0$ which implies that  ${\dot e}_{p}={\dot h}_{d}>0$. For isothermal case, at stationary state (approaching equilibrium), ${\dot e}_{p}={\dot h}_{d}=0$. 

Moreover, for the case where the probability distribution is  either periodic or  vanishes  at the boundary, Tome $et$. $at.$ \cite{ta1} derived the expressions for the entropy production and entropy extraction rates for isothermal case.  Following their approach,   let us rewrite Eq. (3) as 
\begin{eqnarray}
{\partial P\over \partial t}&=& k+{\partial J' \over \partial v} 
\end{eqnarray}
where 
\begin{eqnarray}
k=v{\partial P \over \partial x} +{1\over m}(U'){\partial P \over \partial v}
\end{eqnarray}
and 
\begin{eqnarray}
J'= -{\gamma\over m}vP-{ T(x)\over m^2}{\partial P\over \partial v}.
\end{eqnarray}
The expression $k$ vanishes after imposing a boundary condition.
After some algebra one gets 
\begin{eqnarray}
{\dot e}_{p}&=& -\int {m^{2}J'^{2} \over P T(x) \gamma} dx dv
\end{eqnarray}
and 
\begin{eqnarray}
{\dot h}_{d} &=& -\int  {mvJ' \over T(x) } dx dv
\end{eqnarray} respectively.
In the next sections  we  show that indeed  Eqs. (8)  and  (16)  as well Eqs. (9) and (15) agree  as long as a periodic boundary condition is imposed.  

In general,  since     the  expressions for 
${\dot S}(t)$, ${\dot e}_{p}(t)$  and ${\dot h}_{d}(t)$ can obtained at any time $t$, the analytic expressions for  the change in entropy production,  heat dissipation  and  total entropy can be found analytically via   
\begin{eqnarray}
\Delta h_d(t)&=& \int_{0}^{t}{\dot h}_{d}(t)dt \nonumber \\
\Delta e_{p}(t)&=& \int_{0}^{t}  {\dot e}_{p}(t)  dt \nonumber \\
\Delta S(t) &=&\int_{0}^{t} {\dot S}(t)dt 
\end{eqnarray}
where $\Delta S(t)=\Delta e_p(t)-\Delta h_d(t)$.

{\it  Derivation for the free energy \textemdash} 
The free energy dissipation rate ${\dot F}(t)$   can  be expressed in terms of  ${\dot E}_{p}(t)$   and  ${\dot H}_{d}(t)$. ${\dot E}_{p}(t)$   and  ${\dot H}_{d}(t)$ are the terms that are associated with ${\dot e}_{p}(t)$   and  ${\dot h}_{d}(t)$.   Let us 
now introduce  ${\dot H}_{d}(t)$ for the model system we considered.   The heat dissipation rate   is either  given by Eq. (7) (for any cases) or if a periodic boundary condition is imposed, ${\dot H}_{d}(t)$ is given by 
\begin{eqnarray}
{\dot H}_{d} &=& -\int  {mvJ' dx dv}.
\end{eqnarray}
Equation (18) is notably different from Eqs. (8) and (16), due to the  the term $T(x)$. 
On the other hand, the term  ${\dot E}_{p}$ is  related to  ${\dot e}_{p}$ and it is given by 
\begin{eqnarray}
{\dot E}_{p}&=& -\int {m^{2} J'^{2} \over P \gamma} dx dv.
\end{eqnarray}
The new entropy balance equation  
\begin{eqnarray}
{d S^T(t)\over dt}&=&{\dot E}_{p}-{\dot H}_{d}
\end{eqnarray}
is associated to Eq. (6)  except the term $T(x)$.
Once again, because   the  expressions for 
${\dot S}^T(t)$, ${\dot E}_{p}(t)$  and ${\dot H}_{d}(t)$ can be obtained as a function of  time $t$, the analytic expressions for  the change related to the rate of  entropy production,  heat dissipation  and  total entropy can be found analytically via  
\begin{eqnarray}
\Delta H_d(t)&=& \int_{0}^{t}{\dot H}_{d}(t)dt  \nonumber \\
\Delta E_{p}(t)&=& \int_{0}^{t} {\dot E}_{p}(t) dt  \nonumber \\
\Delta S(t)^T &=&\int_{0}^{t} {\dot S}(t)^{T}dt
\end{eqnarray}
where $\Delta S(t)^T=\Delta E_p(t)-\Delta H_d(t)$.

On the other hand, the internal energy is given by 
\begin{eqnarray}
{\dot E}_{in} = \int ({\dot K}+ v U'_{s}(x))P(x,v,t)dvdx.
\end{eqnarray}
where   ${\dot K}=m{vdv\over dt}$  and $U'_{s}$ denote the rate of kinetic and   potential energy, respectively.
For a Brownian particle that   operates  due to the spatially varying temperature case, the total work done is then given by 
\begin{eqnarray}
{\dot W}&=& \int v f P(x,v,t)dvdx.
\end{eqnarray}
The first law of thermodynamics  can be written as  
\begin{eqnarray}
{\dot E}_{in} = -{\dot H}_{d}(t)-{\dot W}.
\end{eqnarray}
 The change in the internal energy   reduces to  
$
\Delta E_{in}= -\int_{0}^{t}( {\dot H}_{d}(t)+{\dot  W}) dt 
$

As discussed in the work \cite{mu17,muu17,muuu17}, the rate of free energy  is given by ${\dot F}={\dot E}-T{\dot S}$ for isothermal case and  ${\dot F}={\dot E}-{\dot S}^T$ for nonisothermal case  where ${\dot S}^T={\dot E}_{p}-{\dot H}_{d}$. Hence  
we write the  free energy dissipation rate as
\begin{eqnarray}
{\dot F}&=&{\dot E}_{in}- {\dot S}^T \nonumber \\
&=&{\dot E}_{in}-{\dot E}_{p}+{\dot H}_{d}.
\end{eqnarray}
The change in the free energy is given by  
\begin{eqnarray}
\Delta F(t)&=&-\int_{0}^{t} \left(  {\dot W}+ {\dot E}_{p}(t)   \right)dt.
\end{eqnarray} 
For isothermal case, at  quasistatic limit where the velocity  approaches  zero  $ v=0$, ${\dot E}_{p}(t) =0$ and ${\dot H}_{d}(t) =0$ and far from quasistatic limit 
$E_{p}={\dot H}_{d}>0$  which is  expected as   the particle  operates irreversibly.

\section{ Isothermal case}
In this section we discuss the thermodynamic properties for a Brownian particle moving freely without any boundary condition  in underdamped medium  under the influence of   a force $f$ in the absence of a  potential $U'_{s}$. 
The general  expression  for   
the probability distribution $ P(v,t) $   is calculated as 
\begin{eqnarray}
P(v,t)&=&{e^{-{m({-(1-e^{-{\gamma t \over m}})f\over \gamma }+v)^2\over 2(1-e^{-{2\gamma t \over m}})T}}\sqrt{{m\over (1-e^{-{2\gamma t \over m})}t}}\over \sqrt{ 2\pi}}.
\end{eqnarray}
 The average velocity  has a form 
\begin{eqnarray}
\left\langle v(t)\right\rangle=\left({1-e^{-{\gamma t\over m}}\over \gamma}\right)f.
\end{eqnarray}
At steady state ( the long time limit), the velocity  approach $v=f/\gamma$ as expected.

\subsection{ Free particle diffusion  }

For a Brownian particle that moves  in underdamped  medium  without an external force, $f=0$, next let us explore  how the entropy production and  extraction rates behave. From now on, whenever we plot any figures,  we use    the following dimensionless 
load ${\bar f}=fL_{0}/T_{c}$, temperature ${\bar T}=T /T_{c}$ where $T_c$ is the reference temperature of the isothermal medium.
We  also introduced  dimensionless parameter  ${\bar x}=x/L_{0}$.  Hereafter the bar will be dropped.
From now on all  the  figures in this section will be plotted in terms of the dimensionless parameters.

The expression for the  entropy can be readily calculated  by substituting  Eq. (27) in Eq. (4). Figure 1  exhibits that the entropy $S(t)$  increases with time and  saturates to a constant value which  agrees with the results shown in the works \cite{muu17,muuu17}. On the other hand  the entropy production and extraction rates   explored via Eqs. (6), (8) and (9) (see Fig.2).
 The plot  ${\dot e}_{p}(t)$ (red solid line)   and ${\dot h}_{d}(t)$ (black  solid line) as a function of $t$ for parameter choice $\tau=1$  is depicted in Fig. 2. The  figure exhibits that ${\dot e}_{p}(t)$ decreases  as time increases and in long time limit, it approaches its stationary  value ${\dot e}_{p}(t)=0$. On the other hand ${\dot h}_{d}(t)=0$ regardless of $t$.   In the limit $t\to \infty$, ${d S(t)\over dt}=0$ since  ${\dot e}_{p}(t)= {\dot h}_{d}(t)=0$  in the long time limit.

\begin{figure}[ht]
\centering
{
    \includegraphics[width=4cm]{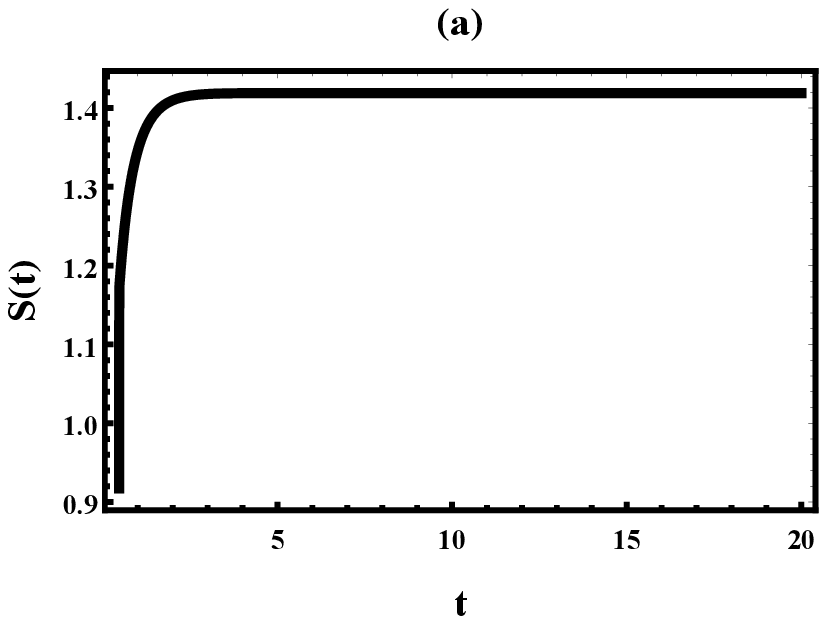}}
\hspace{1cm}
{
    \includegraphics[width=4cm]{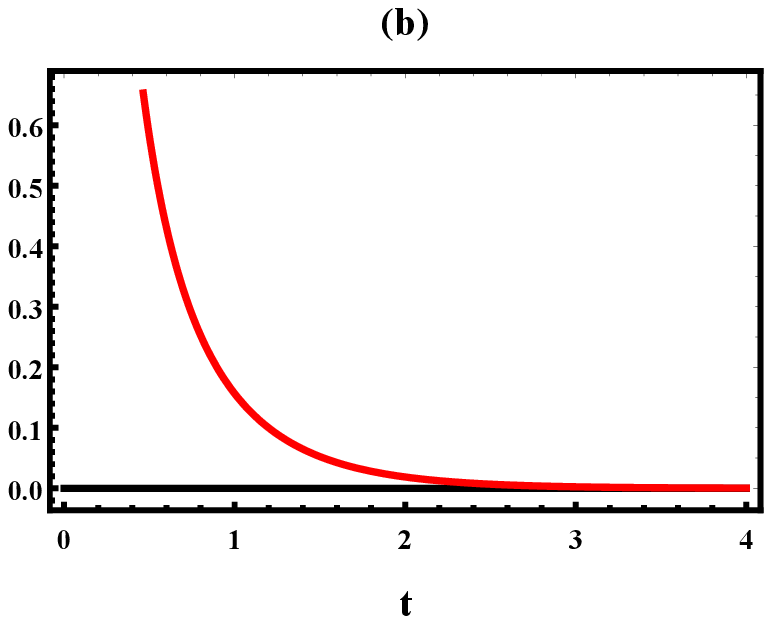}
}
\caption{ (Color online) (a) The entropy $S(t)$ as a function of $t$ for fixed  $\tau=1$. $S(t)$ monotonously  increases with $t$ and saturates to a constant value as $t$ further increases.  
 (b)  The plot  ${\dot e}_{p}(t)$ (black solid line)   and ${\dot h}_{d}(t)$ (red solid line) as a function of $t$ for parameter choice $T=1$. The  figure exhibits that ${\dot e}_{p}(t)$ decreases  as time increases and in the long time limit, it approaches its stationary  value ${\dot e}_{p}(t)= {\dot h}_{d}(t)=0$. 
} 
\label{fig:sub} 
\end{figure}

\subsection{ Particle diffusion in the presence of force}

In the presence of non-zero force, the particle diffuses  under  the influence of the external load. Exploiting  Eqs. (6), (8) and (9),   the dependence of $S(t)$,  ${\dot S}(t)$, ${\dot e}_{p}(t)$  and ${\dot h}_{d}(t)$  on model parameters is explored. In Fig. 2a,  ${\dot S}(t)$ as a function of $t$ is depicted  for fixed values of $T=1$ and $f=1.0$. The figure shows that ${\dot S}(t)$ monotonously  decreases  with $t$ and in the limit $t \to \infty$,  ${ S}(t)$ saturates to zero . 
 Fig. 2b shows   the plot  ${\dot e}_{p}(t)$ as a function of $t$ (red solid lines). In the same figure, the  plot  of ${\dot h}_{d}(t)$  versus  $t$  is shown (black solid  line). The figure exhibits that in the presence of load, ${\dot e}_{p}(t)$ increases  as time increases and in long time limit, it approaches its steady state value  (see the red solid line). ${\dot h}_{d}(t)$   also approaches  its steady state value (see the black solid line) and at steady state 
${\dot h}_{d}(t)={\dot e}_{p}(t)>0$. This also indicates that in the presence of symmetry breaking fields such as external force, the system is driven out of equilibrium.

\begin{figure}[ht]
\centering
{
    \includegraphics[width=6cm]{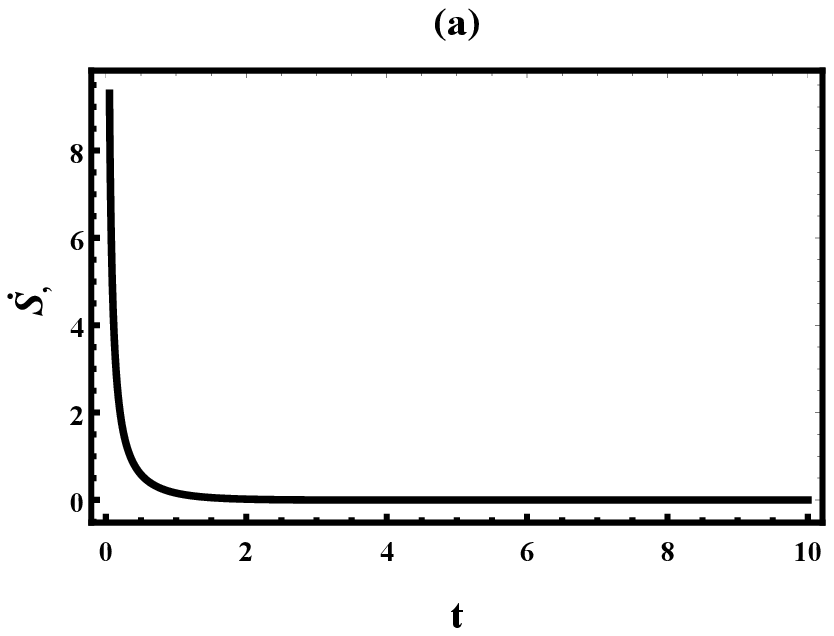}}
\hspace{1cm}
{
    \includegraphics[width=6cm]{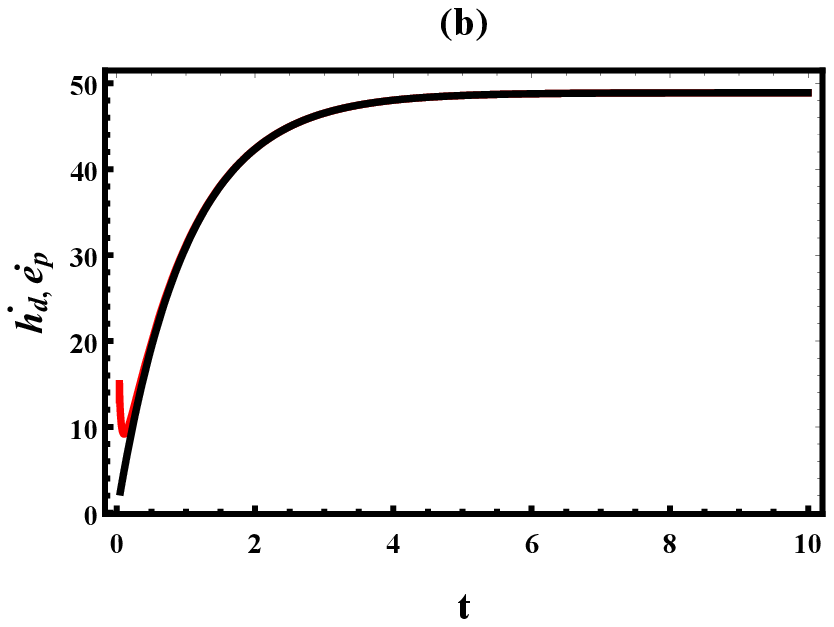}
}
\caption{ (Color online) (a) The entropy $S(t)$ as a function of $t$ for fixed  $\tau=1$. $S(t)$ monotonously  increases with $t$ and saturates to a constant value as $t$ further increases.  
 (b)  The plot  ${\dot e}_{p}(t)$ (red solid line)   and ${\dot h}_{d}(t)$ (black solid line) as a function of $t$ for parameter choice $\tau=1$. The  figure exhibits that ${\dot e}_{p}(t)$ increases  as time increases and in long time limit, it approaches its stationary  value ${\dot e}_{p}(t)= {\dot h}_{d}(t)>0$. 
} 
\label{fig:sub} 
\end{figure}
\begin{figure}[ht]
\centering
{
    \includegraphics[width=6cm]{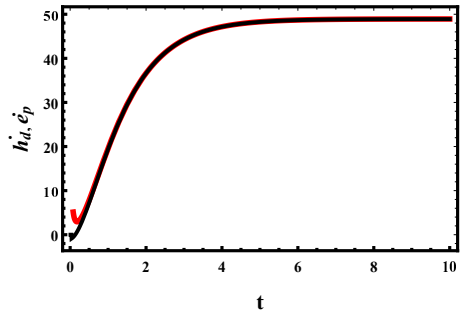}}
\caption{(Color online)   The plot  ${\dot e}_{p}(t)$ (black solid line)   and ${\dot h}_{d}(t)$ (red solid line) as a function of $t$ for parameter choice $\tau=1$. The  figure exhibits that ${\dot e}_{p}(t)$ increases  as time increases and in long time limit, it approaches its stationary  value ${\dot e}_{p}(t)= {\dot h}_{d}(t)>0$.  } 
\end{figure}

Even if no periodic boundary condition is imposed, 
the results shown in Fig. 2 can be also reproduced by employing Eqs. (6), (15) and (16). In fact  Fig. 3 is identical to Fig. 2 except that Fig. 3  is  plotted via Eqs. (6), (15) and (16) while in plotting Fig. 2,  Eqs. (6), (8) and (9)  are used.
 Our analysis  also indicates that the free energy dissipation rate $
{\dot F}$  is always  less than zero $ {\dot F}<0$. As time steps up,  it increases with time and approaches zero in the long time limit.  
All of the  results shown in this work also agree 
  with our previous results \cite{mu17,muu17, muuu17}. As before
$\Delta h_d(t)=h_d(t)-h_d(t_0)>0$,  $\Delta S(t)=S(t)-S(t_0)>0$  or $\Delta e_p(t)=e_p(t)-e_p(t_0)>0$.  

\section{ Nonisothermal case}

\subsection{Periodic boundary condition}
Now let us consider  an important model system where a colloidal particle  that  undergoes a biased random walk in  a spatially varying thermal arrangement in the  presence of  external load  $f$ with no potential. The load 
   is   also coupled  with   a  heat bath  that  
  decreases  from $T_{h}$ at $x=0$  to $T_{c}$ at $x=L_0$  along the  reaction coordinate   in the manner
 \begin{equation} 
 T(x)=\left\{{x(T_{c}-T_{h})\over L_0}+T_{h}\right\}.
   \end{equation}
Here  $L_0$ denotes  the width of the ratchet potential.  $T_{h}$ and $T_{c}$ denote the temperature of the hot and cold baths.

Solving Eq. (3)  at steady state and imposing a periodic boundary condition, the general expression for the probability distribution is obtained as  
\begin{equation} 
P(x,v)=e^{-{L_0 m(f-\gamma v)^2 \over {2 \gamma^2 L_0 (L_0 T_h+(T_c-T_h)
x)}}}   \sqrt{\frac{L_0 m}{2 L_0 \pi  T_h+2 \pi  T_c x-2 \pi  T_h x}}.
   \end{equation}
 The average velocity is found to be  
\begin{equation} 
v={f\over \gamma }.
\end{equation}
 In the absence of force, the velocity approach zero.

Employing Eqs. (6), (8) and (9),  the entropy production and extraction rates  are calculated as 
 \begin{eqnarray} 
{\dot h}_{d}(t)&=&{\dot e}_{p}(t) \nonumber \\
&=&{(2 f L_0)^2  Log[T_c/T_h]\over {4 \gamma
L_0 (T_c-T_h)}}
   \end{eqnarray}
We reproduce  the above result  (using Tome $et$. $at.$ \cite{ta1} approach)   via Eqs. (6), (15) and (16)  as 
 \begin{eqnarray}  
{\dot h}_{d}(t)&=&{\dot e}_{p}(t) \nonumber \\ 
&=&{(2 f L_0)^2 Log[T_c/T_h]\over {4 \gamma
L_0 (T_c-T_h)}}.
   \end{eqnarray}
	Surprisingly, in the limit where the load approaches the the stall  force, ${\dot h}_{d}(t)={\dot e}_{p}(t)=0$. 
	
	The rate of heat dissipation 
is calculated  using  Eq. (7)  (or  Eq. (18)) and it converges to
 \begin{eqnarray} 
{\dot H}_{d}(t)&=&{\dot E}_{p}(t) \nonumber \\  &=&{(f L_0)^2\over  \gamma L_0}.
   \end{eqnarray}
	In the limit where the load approach  zero,
	${\dot H}_{d}(t)={\dot E}_{p}(t)=0$ showing that at quasistatic limt the system  is reversible.
	On the other hand, the rate of work done is given by 
\begin{eqnarray} 
{\dot W}(t)&=&{\dot E}_{p}(t) \nonumber \\ &=&{( f L_0)^2\over  \gamma L_0}.
   \end{eqnarray}
	For isothermal case $T_{h}=T_{c}$ one gets $v=f/\gamma$, ${\dot h}_{d}(t)={\dot e}_{p}(t)=f^2L_{0}/\gamma T_{c}$ and ${\dot H}_{d}(t)={\dot E}_{p}(t)=f^2L_{0}/\gamma$.

All of the results shown in this section  are justified  via numerical simulations by integrating the Langevin equation (1) (employing Brownian
dynamics simulation).  In the simulation, a Brownian particle is initially situated in one of 
the potential wells. Then the trajectories for the particle   is simulated by considering different
time steps $\Delta t$ and time length $t_{max}$. In order to ensure the numerical accuracy $10^{9}$
ensemble averages have been obtained.  Fig. 4 depicts  the plot  $v$  as a function of load $f$. The figure shows the velocity steps up linearly with the load $f$.The simulation results obtained   agree with analytic results. 
	
The plot  ${\dot e}_{p}(t)$   and ${\dot h}_{d}(t)$  as a function of $f$  is depicted in Fig. 5a for parameter choice $\tau=2$.  The figures show that ${\dot e}_{p}(t)$   and ${\dot h}_{d}(t)$ have a nonlinear dependence on the load. Figure 5b   also exhibits  the plot  ${\dot e}_{p}(t)$   and ${\dot h}_{d}(t)$  as a function of $\tau$ for fixed $f=2$. The figure depicts that ${\dot e}_{p}(t)$   and ${\dot h}_{d}(t)$ decrease as the temperature  increases. 

\begin{figure}[ht]
\centering
{
    \includegraphics[width=6cm]{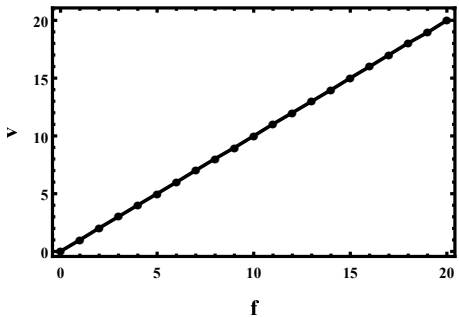}}
\caption{(Color online)   The plot  $v$  as a function of load $f$. The dotted lines are plotted via Brownian dynamic simulation while the solid lines are potted using the analytic Eq. (32).   } 
\end{figure}

\begin{figure}[ht]
\centering
{
    \includegraphics[width=6cm]{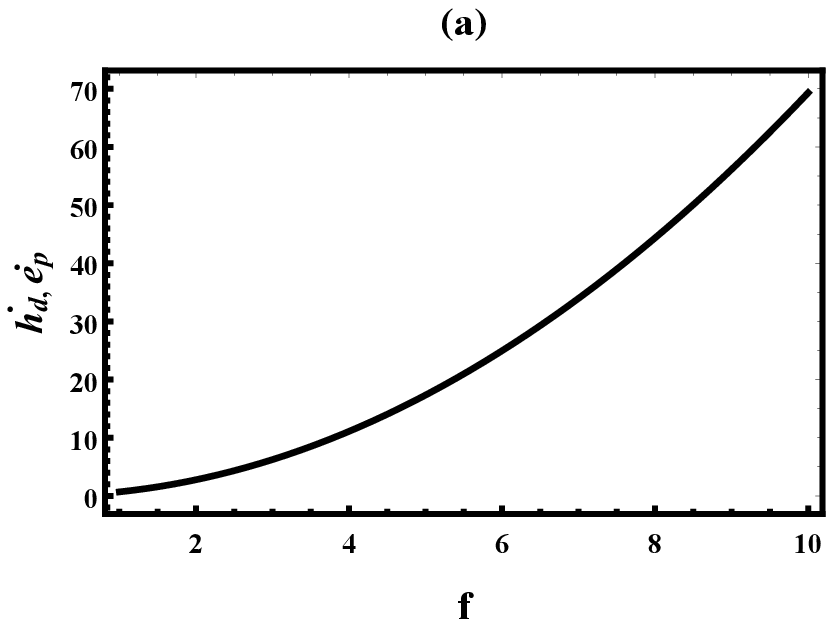}}
\hspace{1cm}
{
    \includegraphics[width=6cm]{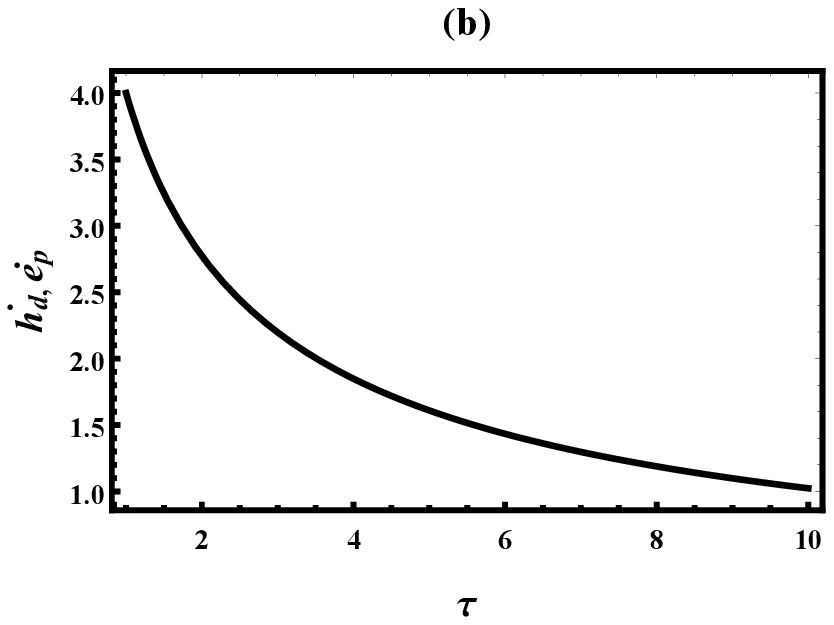}
}
\caption{ (Color online) (a) The plot  ${\dot e}_{p}(t)$   and ${\dot h}_{d}(t)$  as a function of $f$ for parameter choice $\tau=2$. 
 (b)  The plot  ${\dot e}_{p}(t)$    and ${\dot h}_{d}(t)$  as a function of $\tau$ for parameter choice $f=2$. 
} 
\label{fig:sub} 
\end{figure}

\subsection{Nonisothermal case without boundary condition}

All the discussed thermodynamic quantities are quite sensitive to  the choice of the boundary condition. For instance, when no boundary condition is imposed, we find the velocity  for underdamped case as  
 \begin{equation} 
v={2fL_{0}+(T_c-T_h)\over 2\gamma L_0}.
   \end{equation}
	showing the particle stalls when 
	 \begin{equation} 
f_s= {(T_h-T_c)\over 2\gamma L_0}.
   \end{equation}
	When $f<f_s$, the particle velocity $v>0$ and if $f>f_s$, the particle velocity $v<0$. At stall force $f=f_s$, $v=0$. 
The entropy production and extraction rates are given as  
	\begin{eqnarray}
{\dot h}_{d}(t)&=&{\dot e}_{p}(t)	\nonumber \\ 
&=&{(2 f L_0+T_c-T_h)^2 Log[T_c/T_h]\over {4 \gamma
L_0 (T_c-T_h)}}	
\end{eqnarray}
	while 
		\begin{eqnarray}
{\dot H}_{d}(t)&=&{\dot E}_{p}(t)={(2 f L_0+T_c-T_h)^2\over 4 \gamma L_0}.
\end{eqnarray}
Exploiting Eq. (38), one can see that in the limit $f\to f_s$, ${\dot h}_{d}(t) \to 0$ and ${\dot e}_{p}(t)	\to 0$. All of these results indicate that  in the absence of  boundary conditions,  most of the thermodynamic quantities have a functional dependence on  $\Delta T=Th-T_c$ which agrees with the work by Matsuo $et.$. $al.$ \cite{mi1}.

\section{ Brownian particle walking in a ratchet potential where the potential is coupled with a spatially varying temperature}

\begin{figure} 
\epsfig{file=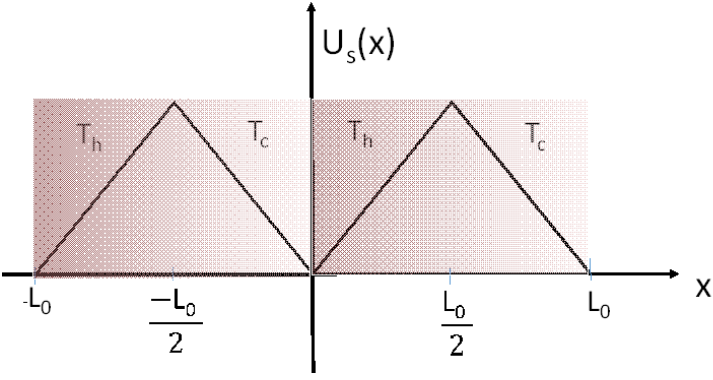,width=10cm}
\caption{ Schematic diagram for a Brownian particle in a piecewise linear  potential in the absence of external load. Due to  the thermal background kicks, the particle ultimately attains a steady state current (velocity) as long  a distinct temperature difference between the hot and the cold reservoirs is retained.}
\end{figure}

In this section, let us  consider  a  Brownian particle  that moves    along the potential $U(x)=U_{s}(x)+fx$  where $f$ and $U_{s}(x)$ denote the load  and  ratchet potential, respectively.  The ratchet potential $U_{s}(x)$ 
 \begin{equation} 
  U_{s}(x)=\left\{\begin{array}{cl}
   2U_{0}[{x\over L_0}],&if~ 0< x \le L_0/2;\\
   2U_{0}[{-x\over L_0}+1],&if~ L_0/2 < x \le L_0;\end{array}\right.
   \end{equation}
   is  coupled  with   a  heat bath  that  
  decreases  from $T_{h}$ at $x=0$  to $T_{c}$ at $x=L_0$  along the  reaction coordinate   in the manner
 \begin{equation} 
 T(x)=\left\{{x(T_{c}-T_{h})\over L_0}+T_{h}\right\}.
   \end{equation}
Here $U_{0}$  denots  the barrier height.  The ratchet  potential has 
 a potential maxima  at  $x=L_0/2$ and potential minima at $x=0$ and $x=L_0$. 
The potential profile repeats itself such that $U_{s}(x+L_0)=U_{s}(x)$.  Next  let us consider the undrdamped case.

\subsection{Underdamped case}

Let us now  explore the dependence of  thermodynamic quantities via numerical simulations by integrating Eq. (1).  In Fig. 7a, the plot of   $v$ as a function of $U_{0}$ is depicted  for fixed $\tau=2$, $f=0.0$, $m=1$ and $\gamma=1$. The figure shows that the velocity peaks at a certain $U_{0}$. On the other hand  the plot of   $v$ as a function of $U_{0}$ is shown in Fig. 7b for fixed $\tau=2$, $f=0.5$, $m=1$ and $\gamma=1$.  The figure  shows that the velocity is negative  below a certain $U_{0}$. As $U_{0}$  steps up the velocity steps up and attains an optimum value.	
\begin{figure}[ht]
\centering
{
    \includegraphics[width=6cm]{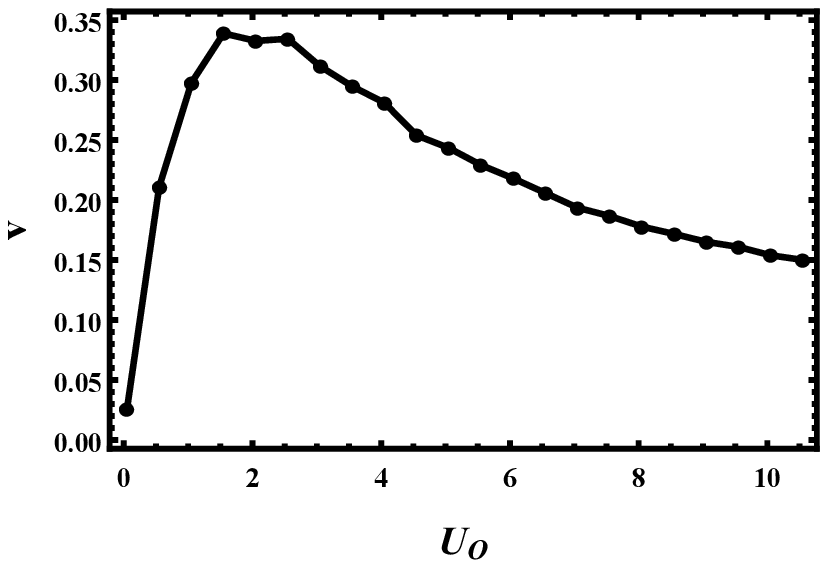}}
\hspace{1cm}
{
    \includegraphics[width=6cm]{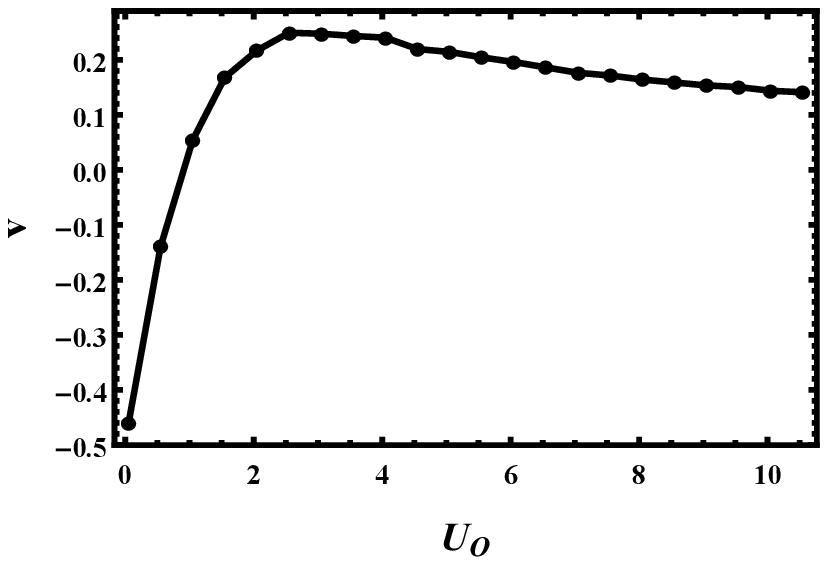}
}
\caption{ (Color online)(a) $v$ as a function of $U_{0}$ for fixed $\tau=2$, $f=0.0$, $m=1$ and $\gamma=1$. 
 (b)  $v$ as a function of $U_{0}$ for fixed $\tau=2$, $f=0.5$, $m=1$ and $\gamma=1$. 
} 
\label{fig:sub} 
\end{figure}

\begin{figure}[ht]
\centering
{
    \includegraphics[width=6cm]{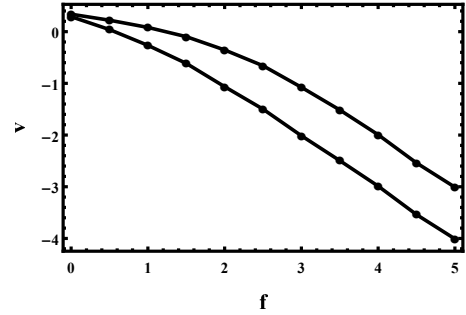}}
\hspace{1cm}
\caption{ (Color online)(a) $v$ as a function of $f$ for fixed $\tau=2$, $U_O=2.0$ (top) and $U_O=1.0$ (bottom), $m=1$ and $\gamma=1$. 
} 
\label{fig:sub} 
\end{figure}
The velocity  decreases monotonously  when the load steps as shown in Fig. 8.  At stall force the velocity becomes zero. Further increases in the load leads to a current reversal.  The plot  ${\dot e}_{p}(t)$   and ${\dot h}_{d}(t)$  as a function of $U_{0}$ for parameter choice $f=2$ is determined via simulations as shown in Fig. 9.

\begin{figure}[ht]
\centering
{
    \includegraphics[width=6cm]{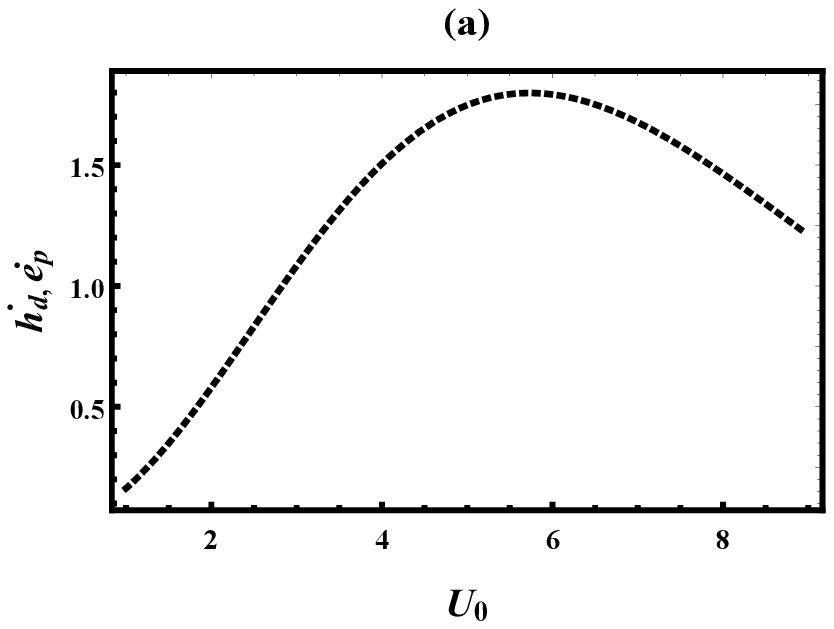}}
\hspace{1cm}
{
    \includegraphics[width=6cm]{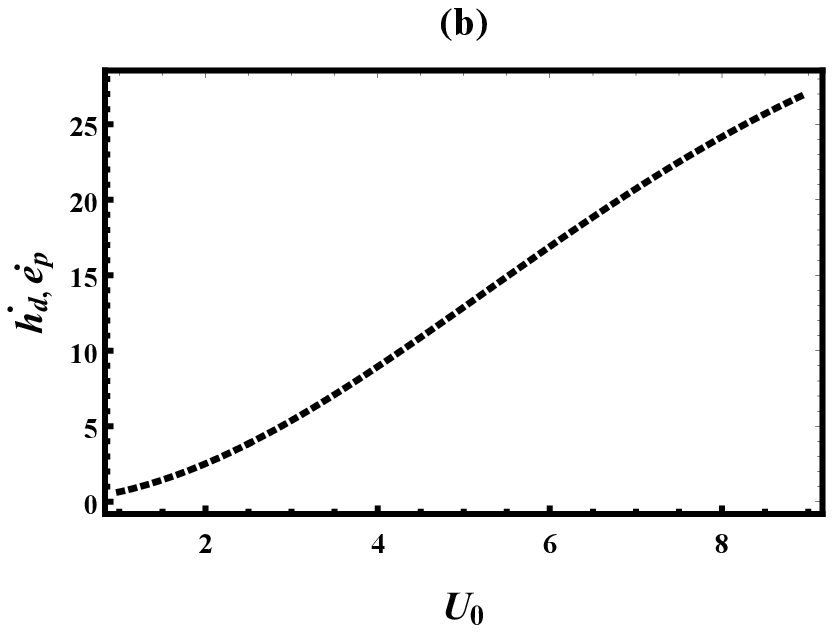}
}
\caption{ (Color online) (a) The plot  ${\dot e}_{p}(t)$   and ${\dot h}_{d}(t)$  as a function of $U_0$ for parameter choice $\tau=12$ and  $f=0.5$. 
 (b)  The plot  ${\dot e}_{p}(t)$    and ${\dot h}_{d}(t)$  as a function of $U_0$ for parameter choice $f=2.0$ (solid line) and  $\tau=2.0$. 
} 
\label{fig:sub} 
\end{figure}

\subsection{Overdamped  case}

In the high friction limit, as discussed before, the dynamics of the particle is governed by the Langevin equation
\begin{eqnarray}
\gamma(x){dx\over dt}&=&-{\partial (U(x) +{T(x)\over 2})\over \partial x} + \sqrt{2k_{B}\gamma(x) T(x)}\xi(t).
\end{eqnarray}  
 The corresponding  Fokker Planck equation is given by 
       \begin{equation} 
       {\partial P(x,t)\over \partial t}={\partial\over  \partial x} (U'(x)P(x,t)+ {T'(x)\over 2}P(x,t)+ T(x){\partial \over \partial x} P(x,t) ) 
   \end{equation}
       where $P(x,t)$ is the probability density of finding the particle at position x
       at time t, $U'(x)={d\over dx}U$.  The  current is given by 
\begin{eqnarray}
J(x,t)&=&-\left[U'(x)P(x,t) +{T'(x)\over 2}P(x,t)+T(x){\partial P(x,t) \over \partial x}\right].
\end{eqnarray}
 In long time limit, the expression for 
the  constant current, $J$, is given in Appendix A.
The change in entropy is given as  \cite{muuu17} 
\begin{eqnarray}
{d S(t)\over dt}&=&{\dot e}_{p}-{\dot h}_{d} \nonumber \\
&=& \int { {J^{2}\over P(x,t)T(x)}+J{ U'(x)\over T(x)}+J{ T'(x)\over 2T(x)}} dx 
\end{eqnarray}
where the entropy production rate ${\dot e}_{p}$ and dissipation rate  ${\dot h}_{d}$ are given as  
\begin{eqnarray}
{\dot e}_{p}&=& \int  {J^{2}\over P(x,t)T(x)}dx 
\end{eqnarray}
and 
\begin{eqnarray}
{\dot h}_{d} &=& \int ( J{ U'(x)\over T(x)}}+{J{ T'(x)\over 2T(x)})dx,
\end{eqnarray} respectively. Here unlike isothermal case, we have additional term $J{ T'(x)\over 2T(x)} dx$.
At steady state ${d S(t)\over dt}=0$ which implies that ${\dot e}_{p}={\dot h}_{d}>0$. At stationary state (approaching equilibrium), $J=0$ since detailed balance condition is preserved. Hence ${\dot e}_{p}={\dot h}_{d}=0$.

In order to relate the free energy dissipation rate with ${\dot E}_{p}(t)$   and  ${\dot H}_{d}(t)$ let us 
now introduce  ${\dot H}_{d}(t)$ for the model system we considered.   The heat dessipation rate is given by 
\begin{eqnarray}
{\dot H}_{d} &=& \int \left(JU'(x)+{JT'(x)\over 2}\right)dx.
\end{eqnarray}
   ${\dot E}_{p}$ is the term related to 
to ${\dot e}_{p}$ and it is given by 
\begin{eqnarray}
{\dot E}_{p}&=& \int \left( {J^{2}\over P(x,t)}\right)dx.
\end{eqnarray}
We have now a new entropy balance equation 
\begin{eqnarray}
{d S(t)^T\over dt}&=&{\dot E}_{p}-{\dot H}_{d} \nonumber \\
&=& \int \left( {J^{2}\over P(x,t)}+JU'(x)+JT'(x) \right). 
\end{eqnarray}
Since the analytic expressions  for $J(x,t)$ and $P(x,t)$ is given in Appendix A, all of  the above  expressions  are  exact but lengthy.  

If one considers a periodic boundary condition  at steady state in the absence of ratchet potential $U_{0}=0$,  the results obtained  are quantitatively   agree with  underdmaped case (Section IV  ) and one gets 
	\begin{eqnarray} 
{\dot h}_{d}(t)&=&{\dot e}_{p}(t) \nonumber \\
&=&{(2 f L_0)^2  Log[T_c/T_h]\over {4 \gamma
L_0 (T_c-T_h)}}
   \end{eqnarray}
	and 
\begin{eqnarray} 
{\dot H}_{d}(t)&=&{\dot E}_{p}(t) \nonumber \\ &=&{(f L_0)^2\over  \gamma L_0}.
   \end{eqnarray}	
				The steady state current is zero at stall load 
 \begin{equation}   
f_{s} = {2U_{0}\over L}{\ln\left[{4\tau \over (1+\tau)^2}\right]\over \ln\left[{1\over \tau}\right]}
  \end{equation}
		which implies the particle velocity $v>0$ when $f<f_s$ and  at stall force $v=0$. When $f>f_s$, $v<0$. 	
 In the quasistatic limit $v \to 0$ ($J\to 0$), the system is reversible.

On the contrary,  in the absence of any boundary condition, the calculated thermodynamic quantities are quantitatively the same as to result shown in Section IV B. For instance, in the absence of potential, the velocity can be calculated  as  $v=JL$. Alternatively, we can also  find $v$ by taking  the time average of Eq. (42) as 
\begin{eqnarray}
\gamma v&=&\left\langle {\partial (fx +{T(x)\over 2})\over \partial x} + \sqrt{2k_{B}\gamma(x) T(x)}\xi(t)\right\rangle \nonumber \\
\gamma v&=&f+{(T_h-T_c)\over 2L_0}\nonumber \\
v&=&{2fL_{0}+(T_c-T_h)\over 2\gamma L_0}.
\end{eqnarray}
Eq. (54) is the same as Eq. (36).  
At this point, we want to stress that  at steady state, most of the derived  physical quantities are similar  both quantitatively and qualitatively  whether the particle is in undrdamped or overdamped medium.

{\it  Derivation for the free energy \textemdash} Next  assuming a periodic boundary condition where the term $T'(x)$ vanishes, let us further  explore the model system.
The expressions for the work done by the Brownian particle as well as the amount heat taken from the hot bath and the  amount of heat given to the cold reservoir  can be derived in terms of the stochastic energetics discussed in the works \cite{am4,am5}.     The heat taken from any heat bath can be evaluated  via \cite{am4,am5}   
${\dot Q}  =\left\langle \left(-\gamma(x){\dot x}+ \sqrt{2k_{B}\gamma(x) T(x)}\right).{\dot x}\right\rangle
  $
while the work done by the Brownian particle against the load is given by 
${\dot W}  =\left\langle f{\dot x}\right\rangle.
$
We can also find the expression for the input heat  $Q_{in}^s$  and $W^s$  as 
\begin{eqnarray}
{\dot Q}_{in} & =&\int_{0}^{L_{0}/2}\left(-\gamma(x){\dot x}+ \sqrt{2k_{B}\gamma(x) T(x)}\right)Jdx \\ \nonumber
&=&\int_{0}^{L_{0}/2}\left[\left({2U_{0}\over L_{0}}\right)+f\right]Jdx \\ \nonumber
&=&U_{0}J+{fL_{0}J\over 2}.
  \end{eqnarray}
Here the integral is evaluated in the interval of $(0,L_{0}/2)$ since the particle has to get a minimal amount of heat input from the heat bath located in the left side of the ratchet potential to surmount  the potential barrier.  The work done is also given by
\begin{eqnarray}
{\dot W} &=&\int_{0}^{L_{0}}fdx=fL_{0}J.
  \end{eqnarray}
	The first law of thermodynamics states that $Q_{in}^s-Q_{out}^s=W^s$  where $Q_{out}^s$  is the heat given to the colder heat bath.  Thus 
	\begin{eqnarray}
{\dot Q}_{out} &=& {\dot Q}_{in}-{\dot W}=U_{0}J-{fL_{0}J\over 2}.
  \end{eqnarray}

	 The  second law of thermodynamics can be rewritten in terms of the housekeeping heat and excess heat.   For the model system we consider, when  the particle  undergoes a cyclic motion, at least it has to get $fLJ$ amount of energy rate from the hot reservoir in order to keep the system at steady state. Hence $ fLJ$ is equivalent to the  housekeeping heat $Q_{hk}$ and we can rewrite Eq. (25)  as 
\begin{eqnarray}
{\dot F} (t)+{\dot E}_{p}(t)={\dot E }_{in}(t)+{\dot H}_{d}(t)=-fLJ=-{\dot Q}_{hk}
\end{eqnarray}
while the expression for the  excess heat ${\dot Q}_{ex}$ is given by
\begin{eqnarray}
Q_{ex}={\dot H}_{d}-{\dot Q}_{hk}.
\end{eqnarray}

For isothermal case, we can rewrite the second law of thermodynamics as \begin{eqnarray}
{\dot S}^T(t)={\dot E}_{p}-{\dot H}_{d}={-\dot F}-{\dot Q}_{ex}
\end{eqnarray}
and 
\begin{eqnarray}
{\dot F}={\dot Q}_{hk}-{\dot E}_{p}.
\end{eqnarray}

At this point we want to stress that such kind of Brownian motor is inherently   irreversible. This can be more appreciated by calculating  the efficiency of the engine. The efficiency is given as
\begin{eqnarray}
\eta= W/Q_{in}.
\end{eqnarray}  
    In the quasistatic limit ($J\to 0$), we find 
\begin{equation}   
\eta = 1-{ln\left[{1+\tau\over 2\tau}\right] \over ln\left[{2\over \tau+1}\right]}
  \end{equation}
  which is approximately equal to the 
 efficiency of the endorevesible heat engine $\eta_{CA}$ 
  \begin{equation}   
\eta_{CA}=1-\sqrt{1/\tau}
\end{equation}   
as long as  the temperature difference between the hot and the cold reservoirs is not large. 
  In order to appreciate this let us Taylor expand Eqs. (63) and (64)  around $\tau=1$ and after some algebra one gets 
  \begin{eqnarray}   
\eta &=& \eta_{CA}={\tau-1\over 2}-{3\over 8}(\tau-1)^2+\ldots \nonumber\\
&=&{\eta_{CAR}\over 2}+{\eta_{CAR}^2\over 8}+{\eta_{CAR}^3\over 96}+\ldots
 \end{eqnarray}
  which exhibits that both efficiencies are equivalent  in this regime. Here $\eta_{CAR}$ is the Carnot efficiency $\eta_{CAR}=1-1/\tau$. 

	Next  we study how  
 the rate of entropy production ${\dot e}_{p}(t)$ and the rate of entropy extraction   ${\dot h}_{d}(t)$ behave. The plot of  ${\dot e}_{p}(t)$  and ${\dot h}_{d}(t)$  as a function of  $f$  is depicted in Fig. 10a   for fixed values of  $U_{0}=2.0$ and   $\tau=12.0$. The plot  ${\dot e}_{p}(t)$    and ${\dot h}_{d}(t)$  as a function of $\tau$ is depicted for parameter choice $f=2.0$ (solid line) and  $U_{0}=2.0$.  Figure 10b indicates that  far from steady state  ${\dot e}_{p}(t)>0$ and ${\dot h}_{d}(t)>0$. 
\begin{figure}[ht]
\centering
{
    \includegraphics[width=6cm]{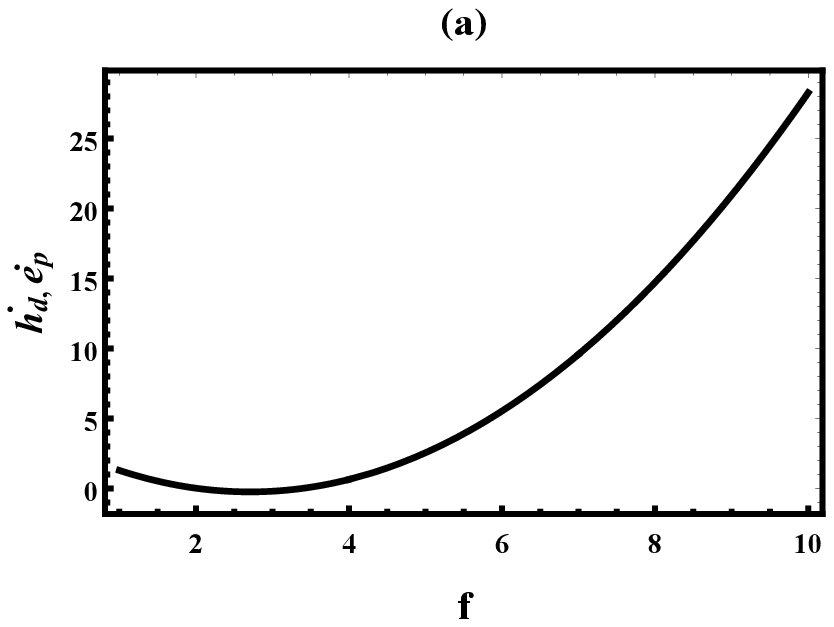}}
\hspace{1cm}
{
    \includegraphics[width=6cm]{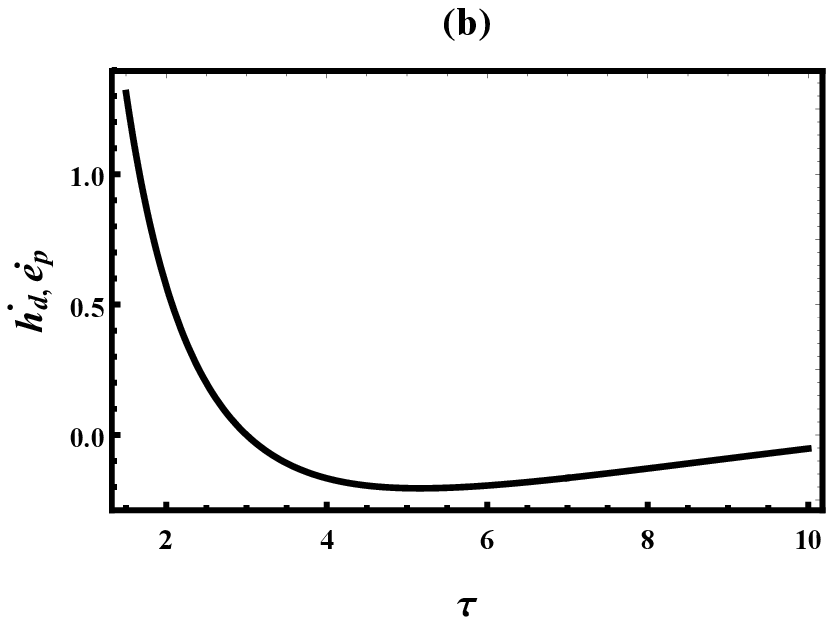}
}
\caption{ (Color online) (a) The plot  ${\dot e}_{p}(t)$   and ${\dot h}_{d}(t)$  as a function of $f$ for parameter choice $\tau=12$ and  $U_{0}=2$. 
 (b)  The plot  ${\dot e}_{p}(t)$    and ${\dot h}_{d}(t)$  as a function of $\tau$ for parameter choice $f=2.0$ (solid line) and  $U_{0}=2.0$. 
} 
\label{fig:sub} 
\end{figure}

\section{ Fluctuation theorem}
As discussed in our previous work \cite{muuu17},  the phase space  trajectory is defined as  $ x(t)={x_{0},x_{1}, x_{\tau}}$ where $x_{s}$ signifies   the phase space at $t=t_{s}$. Whenever the  sequence of noise terms for the total  time of observation  $\xi = {\xi_0,\xi_1,... ,\xi_{s_1}}$ is available, from  the knowledge of the initial point $x_{0}$, $x(t)$ will be then determined.
The probability of  obtaining  the sequence $\xi$ is given as  
 \begin{eqnarray}
P[\xi(t)] \propto e^{[-{1\over 2} \int_{0}^{\tau}\xi^2(t)dt]}.\end{eqnarray}
Since the Jacobian for reverse and forward process is the same, $P[x(t)|x_{0}]$ is proportional 
\begin{eqnarray}
P[x(t)|x_{0}]&\propto & e^{[-{1\over 2} \int_{0}^{\tau}\xi^2(t)dt]} \nonumber \\
&\propto& e^{[-{{1\over 4}\int_{0}^{\tau}dt{(m{dv\over dt}+U+{T\over 2}+{\dot x})^2)\over T}}]}
\end{eqnarray}
Because the Jacobian for reverse and forward process is the same, $P[x(t)|x_{0}]$ is proportional, one gets 
\begin{eqnarray}
{P[x(t)|x_{0}]\over P[\widetilde{x}(t)|\widetilde {x}_{0}]} &=&{ e^{[-{1\over 4}\int_{0}^{\tau}dt{(m{dv\over dt}+U+{T\over 2}+{\dot x})^2\over T}]}\over e^{[-{1\over 4}\int_{0}^{\tau}dt{(m{dv\over dt}+U+{T\over 2}-{\dot x})^2)\over T}]}} \nonumber \\
&=& e^{-\int_{0}^{\tau}dt{( m{dv\over dt}+U+{T\over 2}){\dot x}\over T}]}\nonumber \\
&=&=e^{-\Delta h_d^*(t)}
\end{eqnarray}
Here $h_d^*(t)$ is related  with Eq. (8). 
This implies $ln[{P[x(t)|x_{0}]\over P[\widetilde{x}(t)|\widetilde{x}_{0}]}]=-\Delta h_d^*(t)$. For Markov chain, since $P[x(t)|x(0)] ={P[x(t),x(0)]\over P[x(0)]}$,
$ln[{P[x(t)|x_{0}]\over P[\widetilde{x}(t)|\widetilde{x}_{0}]}]= ln[{P[x(t)]\over P[\widetilde{x}(t)]}]- ln[{P[x_{0}]\over P[\widetilde{x}_{0}]}]=
-\Delta h_d^*(t)$. This also implies that, 
$ln[{P[x(t)]\over P[\widetilde{x}(t)]}]= -\Delta e_h^*(t)$ and  $ln[{P[x_{0}]\over P[\widetilde{x}_{0}]}]=-\Delta s^*(t)$. Clearly   the integral fluctuation relation 
\begin{eqnarray}
\left\langle e^{-\Delta e_h^*(t)}\right\rangle=1.
\end{eqnarray}

\section{Summary and conclusion}

In this work, via Langevin equation  and using Boltzmann-Gibbs nonequilibrium entropy, the general expressions for the free energy, entropy production rate ${\dot e}_{p}$  and    entropy extraction rate ${\dot h}_{d}$  are derived  in terms of velocity and probability distribution considering  underdamped Brownian  motion case.  After extending  the results obtained by Tome. $et.$ $al$ to spatially varying temperature case, we further analyze  our model systems.  We show that 
the entropy production rate ${\dot e}_{p}$ increases  in time and at steady state (in the presence of load), ${\dot e}_{p}={\dot h}_{d}>0$. At stationary state (in the absence of load), ${\dot e}_{p} = {\dot h}_{d}=0$. 
When the particle hops on nonisothermal medium where the medium temperature linearly  decreasing (in the presence of load),  the exact analytic results exhibit that  the velocity approach zero only when the load approach zero. We show that the approximation performed based on Tome. $et.$ $al$  and our   general analytic  expression agree quantitatively. The analytic results also justified via numerical  simulations.

Furthermore,  we   discuss the non-equilibrium thermodynamic features of   a Brownian particle that  hops in a ratchet potential where the potential is coupled with a spatially  varying temperature. It is shown that the operational regime  of such Brownian heat engine is dictated by the magnitude of the external load $f$. The steady state current or equivalently the velocity of the engine is positive   when $f$ is smaller and the engine acts as a heat engine. In this regime ${\dot e}_{p}={\dot h}_{d}>0$.  When $f$ increases, the velocity of the particle decreases and at stall force, we find that ${\dot e}_{p}={\dot h}_{d}=0$ showing that the system is reversible  at this particular choice of parameter.  For large load, the current is negative and the engine acts as a refrigerator. In this region ${\dot e}_{p}={\dot h}_{d}>0$.

In conclusion,   several thermodynamic relations  are derived  for a Brownian particle moving in underdamped medium  by considering different relevant model systems. The present theoretical work not only serves as an important  tool to investigate thermodynamic features of the particle  but also  advances the physics of   nonequilibruim thermodynamics.  

\section*{Appendix A:Derivation of steady state current}
For Brownian particle that moves along the ratchet potential (Eq. (40)) with load $U_{s}(x+L_0)=U_{s}(x)$, in the high friction limit, the dynamics of the particle is governed by the Langevin equation
\begin{eqnarray}
\gamma(x){dx\over dt}&=&-{\partial (U(x) +{T(x)\over 2})\over \partial x} + \sqrt{2k_{B}\gamma(x) T(x)}\xi(t)
\end{eqnarray}  
where $T(x)$ is given in Eq. (41).
 The corresponding  Fokker Planck equation is given by 
       \begin{equation} 
       {\partial P(x,t)\over \partial t}={\partial\over  \partial x} (U'(x)P(x,t)+ {T'(x)\over 2}P(x,t)+ T(x){\partial \over \partial x} P(x,t) ) 
   \end{equation}
       where $P(x,t)$ is the probability density of finding the particle at position x
       at time t, $U'(x)={d\over dx}U$.  The  current is given by 
\begin{eqnarray}
J(x,t)&=&-\left[U'(x)P(x,t) +{T'(x)\over 2}P(x,t)+T(x){\partial P(x,t) \over \partial x}\right].
\end{eqnarray}

The general expression
for the steady state current $J$   for  any periodic potential with or without load is
reported in the works \cite{am14}.  Following the same approach, we find  
the steady state current J  as 
 \begin{equation}   
 J= {-F\over G_{1}G_{2}+HF}.
  \end{equation} 
	where the expressions
for F, G1, G2, and H are given as 
\begin{widetext}
\begin{eqnarray}
F&=& -1+e^{-\frac{2 U_{2} \text{ln}\left[\frac{2}{1+\tau}\right]}{1-\tau}+\frac{2 U_{1} \text{ln}\left[\frac{1+\tau}{2 \tau}\right]}{1-\tau}}, \\
G_1&=&\frac{1-4^{\frac{U_{1}}{1-\tau}} \left(\frac{\tau}{1+\tau}\right)^{\frac{2 U_{1}}{1-\tau}}}{2 U_{1}}+  \nonumber \\ & &
\frac{2^{-1+\frac{2
U_{1}}{1-\tau}} \left(\frac{1+\tau}{\tau}\right)^{-\frac{2 U_{1}}{1-\tau}} \left(-1+4^{\frac{U_{2}}{1-\tau}} \left(\frac{1}{1+\tau}\right)^{\frac{2 U_{2}}{1-\tau}}\right)}{U_{2}},
    \\ 
    G_2&=&\frac{1}{2} \left(\frac{2 \tau}{-1+\tau-2 U_{1}}-\frac{4^{\frac{U_{1}}{-1+\tau}} \left(1+\frac{1}{\tau}\right)^{-\frac{2 U_{1}}{-1+\tau}}
(1+\tau)}{-1+\tau-2 U_{1}}\right) +  \nonumber \\ & &
{1\over 2}\left(\frac{4^{\frac{U_{1}}{-1+\tau}} \left(1+\frac{1}{\tau}\right)^{-\frac{2 U_{1}}{-1+\tau}} \left(1+\tau-2^{1+\frac{2 U_{2}}{-1+\tau}}
\left(\frac{1}{1+\tau}\right)^{\frac{2 U_{2}}{-1+\tau}}\right)}{-1+\tau+2 U_{2}}\right),
\\
   H&=&T_{1}+T_{2}(T_{3}+T_{4}+T_{5}),
    \\
   T_1&=&\frac{\tau \left(-1+4^{\frac{U_{1}}{1-\tau}} \left(\frac{\tau}{1+\tau}\right)^{\frac{2 U_{1}}{1-\tau}}\right)+U_{1}}{2 U_{1}
(1-\tau+2 U_{1})},
\\
    T_{2}&=&2^{-2+\frac{2 (U_{1}+U_{2})}{1-\tau}} \left(\frac{1+\tau}{\tau}\right)^{-\frac{2 U_{1}}{1-\tau}},
     \\
    T_{3}&=&\frac{2^{\frac{1-\tau-2 (U_{1}+U_{2})}{1-\tau}} \left(\frac{1+\tau}{\tau}\right)^{\frac{2 U_{1}}{1-\tau}}}{1-\tau-2 U_{2}}+\frac{2
\tau \left(-4^{-\frac{U_{2}}{1-\tau}}+\left(\frac{1}{1+\tau}\right)^{\frac{2 U_{2}}{1-\tau}}\right)}{(-1+\tau-2 U_{1}) U_{2}},
 \\
T_{4}&=&\frac{2^{-\frac{2 U_{1}}{1-\tau}} (1+\tau) \left(\frac{1+\tau}{\tau}\right)^{\frac{2 U_{1}}{1-\tau}} \left(-2^{-\frac{2 U_{2}}{1-\tau}}+\left(\frac{1}{1+\tau}\right)^{\frac{2
U_{2}}{1-\tau}}\right)}{(1-\tau+2 U_{1}) U_{2}},
 \\
  T_{5}&=&-\frac{2^{-\frac{2 U_{1}}{1-\tau}} (1+\tau) \left(\frac{1+\tau}{\tau}\right)^{\frac{2 U_{1}}{1-\tau}} \left(-2^{-\frac{2 U_{2}}{1-\tau}}+\left(\frac{1}{1+\tau}\right)^{\frac{2
U_{2}}{1-\tau}}\right)}{(1-\tau-2 U_{2}) U_{2}}.
\end{eqnarray}

\end{widetext}
Here  $U_{1}=U_{0}+f/2$  and $U_{2}=U_{0}-f/2$.  The expression for the velocity is then given by $V=LJ$.

\section*{Acknowledgment}
I would like to thank Blaynesh Bezabih and Mulu  Zebene for their
constant encouragement.

\end{document}